\documentclass[%
 aip,
 jmp,%
 amsmath,amssymb,
 preprint,%
author-year,%
]{revtex4-1}

\usepackage{dcolumn}
\usepackage{graphicx,amssymb,color}
\usepackage{gensymb}
\usepackage{bm}

\begin{document}

\begin{center}
\noindent\LARGE{\textbf{A Brief Review on Syntheses, Structures and Applications of Nanoscrolls}}
\vspace{0.6cm}

\noindent\large{\textbf{E. Perim $^a$, L. D. Machado and
D. S. Galvao $^{\ast}$ \textit{$^{a}$}}}\vspace{0.5cm}

\textit{$^{a}$ Instituto de F\'isica ‘Gleb Wataghin’, Universidade Estadual de Campinas, 13083-970, Campinas, SP, Brazil.}

$^{\ast}$ Corresponding author: galvao@ifi.unicamp.br
\end{center}


\begin{abstract}

\section{}
Nanoscrolls are papyrus-like nanostructures which present unique properties due to their open ended morphology. These properties can be exploited in a plethora of technological applications, leading to the design of novel and interesting devices. During the past decade, significant advances in the synthesis and characterization of these structures have been made, but many challenges still remain. In this mini review we provide an overview on their history, experimental synthesis methods, basic properties and application perspectives.

\end{abstract}

\section{Introduction}

\begin{figure}
\begin{center}
\includegraphics[width=15cm]{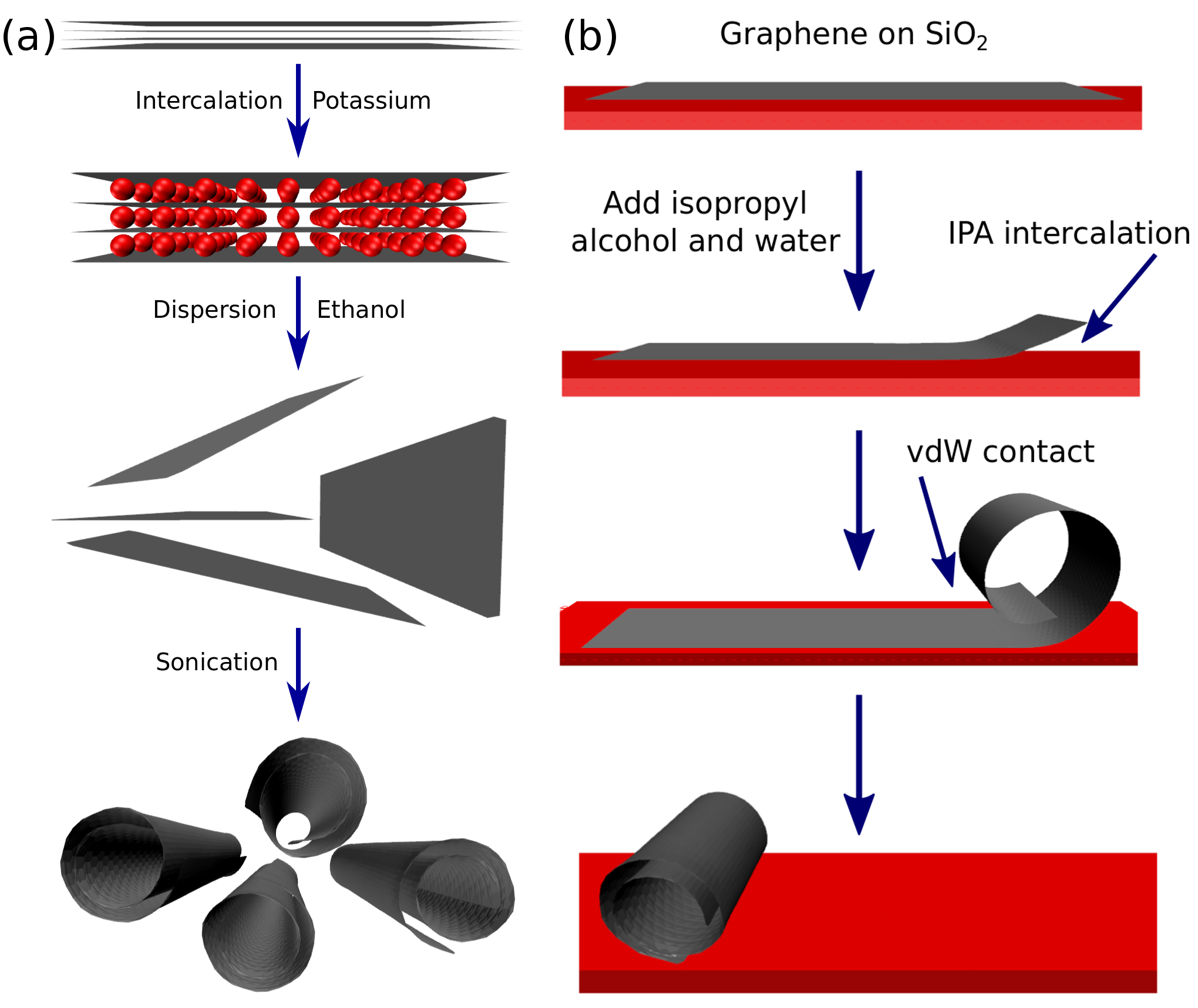}
\end{center}
\textbf{\refstepcounter{figure}\label{fig:01} Figure \arabic{figure}.}{ Scheme of some methods used to synthesize carbon nanoscrolls. (a) Graphite is intercalated with potassium, then dispersed in ethanol and then the mixture is sonicated. (b) Graphene is first mechanically exfoliated from graphite and deposited on SiO$_2$. Then a droplet of isopropyl alcohol (IPA) and water is placed on the monolayer and evaporated. Both methods lead to well-formed nanoscrolls.}
\label{figure1}
\end{figure}

%
%
Nanoscrolls consist of layered structures rolled into a papyrus-like form, as shown in the bottom frame of fig.~\ref{figure1}(a). They are morphologically similar to multiwalled nanotubes, except for the fact that they present open extremities and side, since the edges of the scrolled sheets are not fused. Similarly to nanotubes, nanoscrolls can be constructed from different materials, such as carbon, boron nitride and others. These scrolled structures were first proposed by Bacon in 1960, as a consequence of his observations of the products from arc discharge experiments using graphite electrodes (\cite{bacon1960growth}). Decades later, the advances in imaging techniques made possible the confirmation of his structural model (\cite{dravid1993buckytubes}). Iijima's synthesis of multiwalled carbon nanotubes (MWCNTs) brought renewed attention to nanoscrolls (\cite{iijima1991helical}), as they were later proposed to be part of MWCNTs synthesis (\cite{zhang1993carbon, amelinckx1995structure}). However, only more recently nanoscrolls have started attracting attention as individual and unique structures (\cite{tomanek2002mesoscopic}), as a consequence of the development of efficient methods especially designed towards the production of carbon nanoscrolls (CNSs) (\cite{viculis2003chemical, shioyama2003new}). This motivated a significant number of theoretical and experimental studies on the properties and possible CNS applications.

The characteristic which sets CNSs apart from MWNTs are their unique open ended morphology. These structures are kept from unrolling back into a planar morphology due to van der Waals interactions between overlapping layers (\cite{braga2004structure}). As a consequence, CNSs can have their diameter easily tuned (\cite{shi2010tunable}), can be easily intercalated (\cite{mpourmpakis2007carbon}). They offer wide solvent accessible surface area (\cite{coluci2007prediction}) while sharing some of electronic and mechanical properties with MWCNTs (\cite{zaeri2014elastic}) and preserving the high carrier mobility exhibited by graphene.

\section{Synthesis}

Carbon scrolls were first reported as byproducts of arc discharge experiments using graphite electrodes (\cite{bacon1960growth}). In this kind of experiment the extremely high energies allow the formation of several different carbon structures besides nanoscrolls, such as, nanotubes and fullerenes (\cite{kriitschmer1990solid, ugarte1992morphology, saito1993growth}). However, the high cost, low yield and non-selectivity of this method limits its wide use. The first method designed to produce CNSs at high yield, reaching over 80\%, was developed only decades later (\cite{kaner2002chemical}). This process consists in three consecutive steps. Firstly, high quality graphite is intercalated with potassium metals, then it is exfoliated via a highly exothermal reaction with aqueous solvents. Lastly, the resulting dispersion of graphene sheets are sonicated resulting in CNSs (\cite{viculis2003chemical}) - see fig.~\ref{figure1}(a).  The strong deformations caused by the sonication process leads the solvated sheets to bend and, in case of overlapping layers, to scroll. As calculations pointed out (\cite{braga2004structure}), once significant layer overlap occurs the scrolling process is spontaneous and driven by van der Waals forces. The efficiency of this method has lead it to be adopted in other studies (\cite{roy2008synthesis}). A very similar method was shortly after developed  (\cite{shioyama2003new}), the most significant difference being the absence of sonication. In this case longer times are necessary for graphene sheets in solution to spontaneously scroll .

Both these chemical methods use donor-type intercalation compounds which are highly reactive, demanding the use of inert atmosphere during the process. In order to avoid this limitation, a variation of Viculis et al.'s method was devised utilizing acceptor-type intercalation compounds, namely graphite nitrate, which is much more stable and thus eliminates the need for an inert atmosphere (\cite{savoskin2007carbon}). However, the most significant drawbacks of this chemical approach are the poor morphologies of the resulting nanoscrolls, the inability to control the number of scrolled graphene layers, and also the possibility of defects being introduced during the chemical process. In order to overcome these issues, a new method was later developed, offering higher control over the final product. In this new method, graphite is mechanically exfoliated using the scotch tape method and the extracted graphene layers are then deposited over SiO$_{2}$ substrates. Then a drop of a solution of water and isopropyl alcohol is applied over the structures and the system let to rest for a few minutes. After this, the system is dried out and spontaneously formed CNSs can be observed (\cite{xie2009controlled}) - see fig. ~\ref{figure1}(b). It is believed that surface strain is induced on the graphene layer as a consequence of one side being in contact with the solution and the other being in direct contact with the substrate. Once this strain causes the edges to lift, solvent molecules can occupy the space between layer and substrate, further bending the graphene sheets. As some deformation causes overlap on the layers, the scrolling process becomes spontaneous. While this method offers higher control over produced CNSs, on the other hand it is difficult to scale and more sensitive to defects in the graphene layers.

Synthesis of high quality CNSs from microwave irradiation has also been reported (\cite{zheng2011production}). In this method, graphite flakes are immersed into liquid nitrogen and then heated under microwave radiation for a few seconds. As graphite presents very good microwave absorption, sparks are produced, which are believed to play a key role in the process, as their absence hinders high CNS yields. The resulting product is then sonicated and centrifugated, resulting in well formed CNSs.  

More recently, a purely physical route to CNSs synthesis method was proposed on theoretical grounds (\cite{xia2010fabrication}). In this method a carbon nanotube (CNT) is used to trigger the scrolling of a graphene monolayer. Due to van der Waals interactions, the sheet rolls itself around the CNT in order to lower the surface energy in a spontaneous process. The advantage of this method would be being a dry, non-chemical, room-temperature method. However, it has been shown that the presence of a substrate can significantly affect the efficiency of this method (\cite{zhang2010carbon}). In order to circumvent these limitations, simple changes in substrate morphology have been proposed (\cite{perim2013controlled}). The same principle has been used to propose a method for producing CNS-sheathed Si nanowires (\cite{chu2011fabrication}). However, an experimental realization of such process has yet to be reported. Even more recently, CNSs have been proposed to form from diamond nanowires upon heating (\cite{sorkin2014mechanism}).

Nitrogen-doped graphene oxide nanoscrolls have also been successfully produced (\cite{sharifi2013formation}). Maghemite ($\gamma$-Fe$_{2}$O$_{3}$) particles were used to trigger the scrolling process. In this case, it is believed that the magnetic interaction between the nitrogen defects and maghemite particles is the governing effect in this process. This is supported by the fact that removal of $\gamma$-Fe$_{2}$O$_{3}$ particles causes the scroll to unroll in a reversible process, different from the observed for pure CNSs.

\section{Structure}
\label{struct}

From a topological point of view, scrolls can be considered as sheets rolled up into Archimedean spirals. Hence, the polar equation used to describe these spirals,
\begin{equation}
r=r_0+\frac{h}{2\pi}\phi, 
\label{eq1}
\end{equation}
can be used to determine the points $r$ that belong to the scroll, for a given core radius $r_0$, interlayer spacing $h$, and number of turns $N$ ($\phi$ varies from $0$ to $2\pi N$). See fig.~ \ref{fig:02}(a). In addition, in order to fully determine the geometry of the scroll, the axis around which the scroll was wrapped must be given (see fig.~\ref{fig:02}(b)). Therefore, armchair, zigzag and chiral nanoscrolls exist, although scroll type is not fixed during synthesis and interconversion can occur in mild conditions, due to the open ended topology (\cite{braga2004structure}).

Particular values of $r_0$, $h$ and $N$ for the general scroll geometry described above will depend on the properties of the composing scroll material. As shown by \cite{shi2011constitutive}, the core radius $r_0$ can be determined from the interaction energy between layers ($\gamma$), the bending stiffness of the composing material (D), the interlayer separation (h), the length of the composing sheet (B), and the difference between the inner ($p_i$) and outer pressure ($p_e$), as described by the following equation:
\begin{equation}
p_i-p_e= \frac{D}{2h} \left ( \frac{2\gamma\ h}{D}-\frac{1}{r_0}+\frac{1}{R} \right ) \left ( \frac{1}{r_0}+\frac{1}{R} \right ), 
\label{eq2}
\end{equation}
where $R= \sqrt{\frac{Bh}{\pi}+r_{0}^2}$.

Density Functional Theory (DFT) calculations, carried out without pressure difference, predicted that the minimum stable core diameter is 23 $\AA$ (\cite{chen2007structural}). The interlayer spacing, however, depends mostly on the interaction energy between layers, although several factors can alter its value, like the presence of defects (\cite{tojo2013controlled}). Given a core size and an interlayer distance, the number of turns can be obtained after fully wrapping a given sheet width.

In order to form a nanoscroll, there is an elastic energy cost associated with bending the sheet and a van der Waals (vdW) energy gain associated with the creation of regions of sheet overlap. For graphene, in particular, if the core size and number of turns is appropriate, the vdW energy gain is large enough to make the final scrolled structure even more stable than its initial planar configuration (\cite{braga2004structure}). There is, however, an energy barrier associated with the formation of scrolls, since the initial bending cost is not followed by energy gains. Various theoretical and experimental methods have been devised to overcome this barrier, as discussed in the previous section.

\section{Mechanical Properties}

In this and the next section we will restrict ourselves to the discussion of carbon-based scrolls. In section 6 non-carbon scrolls will be also addressed.

So far we have discussed mainly equilibrium geometry properties, but there are several studies addressing the mechanical response of scrolls to applied agents and/or forces. The first of these studies was carried out by \cite{zhang2012buckling} using molecular mechanics, and studied the response of CNSs to axial compression, twisting and bending. With regard to compression, the authors found that the axial stiffness of CNTs and CNSs with similar diameters and number of layers is about the same, but that nanoscrolls buckle under a significantly smaller strain. The authors argued that the free ends of the CNSs tend to wrinkle and are vulnerable to further buckling, which then propagates inward from the ends. With relation to torsions, the paper reported that both the torsional rigidity and critical strain are much lower for CNSs when compared to CNTs, a result that was attributed again to the open topology. Finally, regarding bending, the authors found about the same response for size-similar CNTs and CNSs. To explain this result, the authors reasoned that bending buckling began in their simulations at localized kinks, and that therefore the global topology of the structure did not matter as much in this case. Also, they reported that increasing the number of turns increases the bending rigidity, but decreases the critical buckling strain. 

\cite{song2013atomic} also used molecular mechanics to study the response of CNSs to compressive stresses. Similarly to the previously discussed work, they also found that compressive buckling started at the free ends. By adding nickel nanoparticles to the ends, they managed to stabilize the dangling bonds, preventing the wrinkling of the edges. This resulted in a slight increase in the elastic modulus (from 950-970 to 1000-1025 GPa) and in a decent increase in the compression strength of the nanoscrolls (from 40-47 to 45-51 GPa). The reason the values above are presented in a range is that the modulus and strength also were found to depend slightly on core radius and on chirality. Also, note that the compression strength corresponds to the point in the stress-strain curve in which increasing the strain leads to a decrease in the stress. The authors also studied the influence of adding a CNT to the inside of the scroll, and found that it did not significantly influence the critical strain value or the deformation morphology.

A third paper on mechanical properties of CNSs, by \cite{zaeri2014elastic}, studied their response under tensile and torsional stresses. Unlike the previous studies, in this one a finite element based approach was used in the description of the elastic properties. Calculations were performed with and without vdW interactions. Regarding the tensile studies, the authors reported a Young's modulus of about 1100/1040 GPa with/without considering vdW interactions. They also studied the influence of changing chirality, core radius, number of turns and scroll length on the Young's modulus, but found only a small dependence for each case. Regarding the application of the torsional stresses without explicitly taking into account vdW interactions, the authors found no shear modulus dependence on chirality and a moderate decrease of its value as the core radius size increased (from 48 to 36 GPa). More importantly, the shear modulus greatly increased as the number of layers increased (from 20 to 100 GPa) and greatly decreased as the length increased (from 95 to 10 GPa). For the first effect, the authors first argued that the inner and outer layers could not resist torsion well due to the open edges, and then explained that the shear modulus increased as the number of torsion resisting intermediate layers increased. To explain the second observation, the authors suggested that longer inner and outer edges increased the weakening effect, though no explanation was given as to why this should happen. Regarding the influence of vdW interactions, it was found that they increased the shear modulus ten times, from about 50 to 500 GPa.

The vibrational properties of CNSs were studied by \cite{shi2009gigahertz}. Using theoretical modeling and molecular dynamics simulations, they showed that the "breathing" (radial) oscillations can described by the interaction energy between layers ($\gamma$), the bending stiffness of the composing material (D), the interlayer separation (h), the length of the composing sheet (B), the density of the material ($\rho$), the internal radius ($r_0$) and the difference between the inner ($p_i$) and outer pressure ($p_e$). The vibrational frequency is described by the following equation:
\begin{equation}
\omega_0= \frac{\pi\ D}{4\rho\ B^3h^2} \left ( \frac{\alpha^3(\alpha+3)}{(\alpha+1)^2}-2\frac{\gamma}{D}\sqrt{\frac{Bh^3}{\pi}} \frac{\alpha^2\sqrt{\alpha}}{(\alpha+1)^{3/2}}-2\frac{Bh^2p}{\pi\ D}\alpha \right ), 
\label{eq3}
\end{equation}
where $\alpha=\frac{Bh}{\pi\ r_{0}^2}$. For a CNS of 10nm length the aforementioned equation leads to a frequency of almost 60GHz.

Much has yet to be done regarding the mechanical properties of carbon nanoscrolls. For instance, the ultimate tensile strength and strain and the fracture pattern of scrolls has yet to be reported. Moreover, both studies regarding the application of compressive stress remark that the results might depend on the scroll length - it is possible that the CNS might bend under compression if they are long enough. One last example is that it remains to be tested whether elements other than nickel could improve the mechanical properties of carbon nanoscrolls.

\section{Applications}

\cite{xie2009controlled} have built a CNS based electronic device in which a nanoscroll was placed between two metallic contacts over a SiO$_2$/Si substrate. The advantage of using a CNS instead of a CNT lies in the ability of nanoscrolls to carry current through all of its layers, while MWCNTs only carry current through the outermost layer, since the inner ones do not make direct contact. It was shown that a CNS was able to withstand a current density up to 5.10$^7$A/cm$^2$, indicating its suitability for circuit interconnects. A detailed theoretical study on the quantum electron transport in CNSs was carried by \cite{li2012quantum}, showing a strong dependence of the conductance on the nanoscroll radius as well as on the temperature.

Another possible application of CNSs is as electroactuators (structures which present mechanical response to adding/removing charges) (\cite{rurali2006prediction}). The authors carried their study by adding/removing charges (up to $\pm0.055 e$/atom) and then performing geometrical optimization using density functional theory (DFT) methods. In the axial direction, the reported length variation for CNSs ($\sim 0.4 \%$) was comparable to those reported for CNTs (\cite{verissimo2003electromechanical}). In the radial direction, however, the authors reported diameter variations of up to 2.5$\%$, a result that is an order of magnitude larger than what has been reported for CNTs. These results can be understood by considering the open geometry of CNSs, that enable large increases of interlayer distances without breaking bonds. Although for low values of charge injection the calculated response depended on whether electrons or holes had been injected, for large enough values the mechanical response was found to be dominated by the electrostatic layer repulsions. The calculations also showed that the extra charge accumulated at the edges and at the central part of the scroll. Since the authors considered little more than one turn of scroll, only charges at the edge contributed to the interlayer expansion, suggesting that even larger expansion could be possible for CNSs with more layers, in which the central charges would also play a role. 
This large radial actuation has been proposed as a method to control the flow rate in nanoscopic water channels, nanofilters and ion channels \cite{shi2010tunable}. Other possible method to produce mechanical response from scrolls is by applying electric fields. \cite{shi2010translational} reported that the application of an eletric field causes a decrease in the interaction between scroll layers, which in turn could be used to controllable roll/unroll CNSs.

\begin{figure}
\begin{center}
\includegraphics[width=15cm]{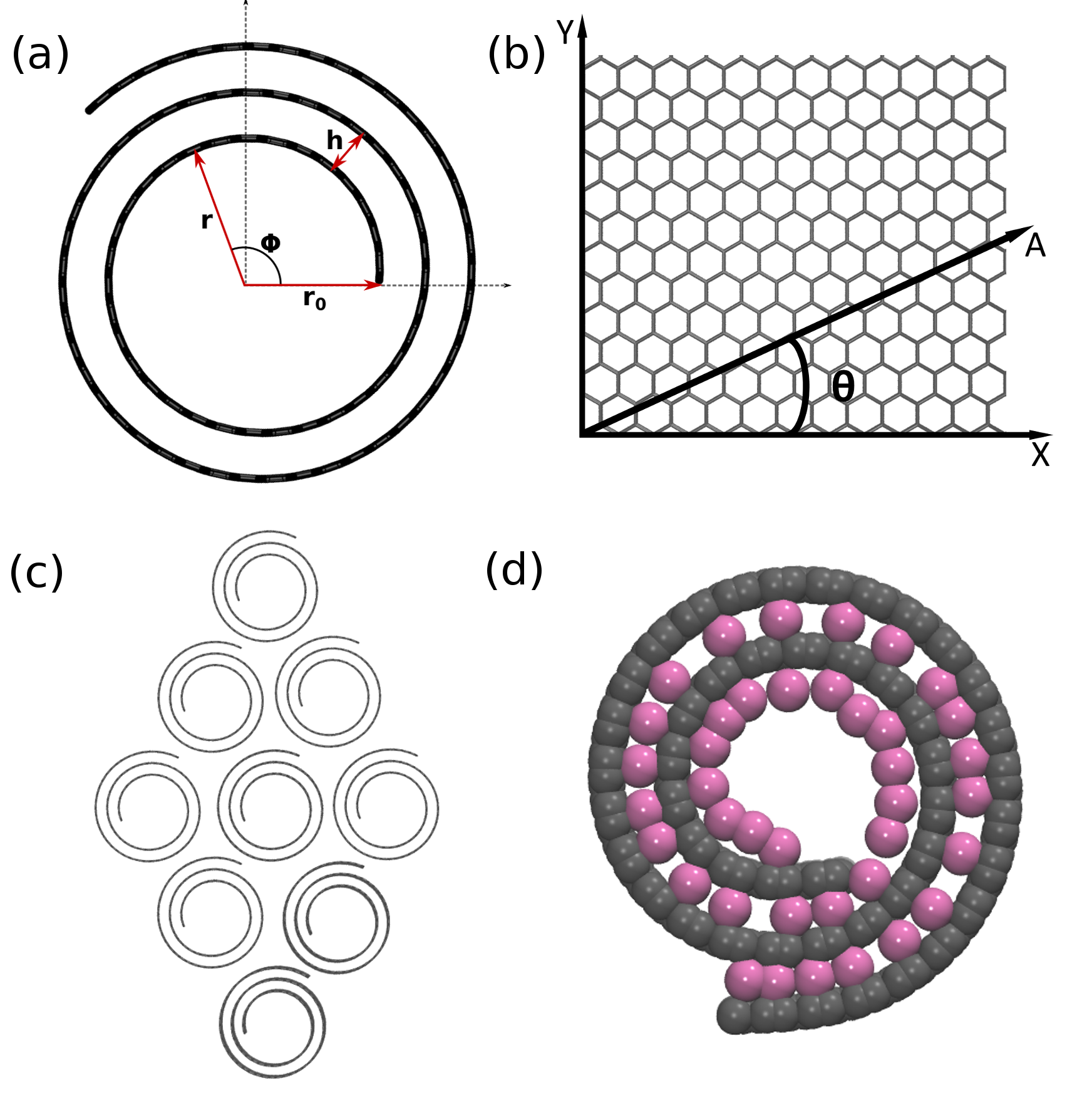}
\end{center}
\textbf{\refstepcounter{figure}\label{fig:02} Figure \arabic{figure}.}{ (a)
Scheme showing the CNS cross-section. See eq. \ref{eq1} and section \ref{struct} for the definition and the relationship between $r_0$, $h$, $r$, and $\phi$. (b)
Scheme for rolling up a CNS from a graphene sheet. $\theta$ is the rolling angle and \emph{A} is the CNS axis. (c) Nanoscroll bundle. (d) Cross-section view of an intercalated nanoscroll. See text for discussions.}
\label{figure2}
\end{figure}

Another application that has received considerable attention is the use of CNSs in gas storage, particularly hydrogen. \cite{coluci2007prediction} were the first to investigate the use of carbon nanoscrolls as a medium to store hydrogen, using classical grand-canonical Monte Carlo simulations. The authors reported that
the interlayer galleries of the scrolls were only available for H$_2$ storage for interlayer spacings larger than 4.4 $\AA$. For instance, at 150 K, the gravimetric storage of CNSs was predicted to increase from 0.9\% to about 2.8\% for crystal packed scrolls and from 1.5\% to 5.5\% for scroll bundles (see fig.~\ref{fig:02}(c)) when the interlayer spacing increased from 3.4 $\AA$ to 6.4 $\AA$. The authors used a fixed pressure of 1 MPa in all their calculations. One possible way to experimentally realize this increase in interlayer spacing is by intercalating scroll layers with alkali atoms (fig.~\ref{fig:02}(d)), and \cite{mpourmpakis2007carbon} reported that this method indeed works well. \cite{coluci2007prediction} also reported that the gravimetric storage decreased greatly by increasing the temperature to 300 K, and \cite{braga2007hydrogen} used molecular dynamics simulations to show that it is possible to cyclically absorb and reabsorb hydrogen from scrolls by 
decreasing and then increasing the temperature. \cite{huang2013molecular} also studied possible mechanisms for delivering hydrogen from the scrolls, and reported that other effective methods include twisting the scrolls and decreasing their interlayer distance. Finally, note that calculations performed by \cite{peng2010computer} reported that CNSs with expanded interlayer distances could also be used to store methane or trap carbon dioxide.
 
It should be noted that recently the use of CNSs in a variety of experimental devices has been gaining momentum. For instance, scrolls have already been used as supercapacitors (\cite{zeng2012supercapacitors,yan2013nanowire}), in batteries (\cite{tojo2013controlled,yan2013nanowire}), in catalysis (\cite{zhao2014facile}) and in sensors (\cite{li2013graphene}). When compared to similar planar graphene based devices, the ones based on CNS were found to present superior performance. The difference was attributed to the open ended carbon nanoscroll topology.

\section{Nanoscrolls From Some Other Materials}

The successful isolation of single layer graphene (\cite{novoselov2004electric}) has created a revolution in carbon-based materials. In part because of this, there is renewed interest in other two-dimensional materials, which has led to some very significant synthesis advances. Hexagonal boron nitride (hBN) (\cite{meyer2009selective, jin2009fabrication}), graphene-like carbon nitride (\cite{li2007synthesis}), graphyne (\cite{kehoe2000carbon, haley2008synthesis}), silicene (\cite{vogt2012silicene}) and, very recently, germanene (\cite{davila2014germanene}) monolayers have been already produced. Each one of these structures present their own unique properties, therefore it is a natural question whether it is possible to use them to form novel nanoscrolls.

Hexagonal boron nitride nanoscrolls (BNNSs) were theoretically predicted some years ago (\cite{perim2009structure}). These structures were predicted to be even more stable than their carbon counterparts, due to stronger interlayer van der Waals interactions, and should present analogous morphology. Thus, BNNSs should present many of the CNSs properties, like easily tunable diameter and very large solvent accessible surface area, as well as, new ones due to the electric insulating and chemically inert hBN nature. Quite recently, these predictions were confirmed as three independent groups reported the successful BNNS synthesis (\cite{li2013exfoliation, suh2014formation,chen2013shear}). The first reported method consists in exposing hBN crystals to an intense solvent flow inside a spinning disc processor. The shear forces are believed to exfoliate hBN layers which are then scrolled, forming BNNSs (\cite{chen2013shear}). However, the yield of this method is considerably low ($\sim$5\%). A more simple method has been proposed, in which molten hydroxides are used to exfoliate hBN crystals (\cite{li2013exfoliation}). A NaOH/KOH mixture was added to hBN and then thoroughly ground, subsequently being heated at 180$^{\circ}$ C. The analysis of the product revealed presence of BNNSs, however, at even lower yields than the previous method. BNNSs have also been produced by the interaction between exfoliated hBN and lithocholic acid (\cite{suh2014formation}), in a self-assembly process. The considerably lower yields of these methods when compared to the ones utilized for producing CNSs indicate the higher difficulty of exfoliating hBN crystals due to stronger interlayer interactions.

Carbon nitride nanoscrolls (CNNSs) have similarly been predicted to be stable (\cite{perim2014novel}). Three different graphene-like carbon nitride structures have been successfully synthesized (\cite{li2007synthesis}) with varying pore sizes and simulations predict that all three of them should be able to form stable nanoscroll structures. The existence of pores in these structures reduce the contact area between overplapping layers, leading to weaker van der Waals interactions and thus less stable nanoscrolls. On the other hand, it also means lower mass density and also easier intercalations, which means these scrolls could be even more suited to hydrogen storage and similar applications. Up to now, no CNNS synthesis has been achieved.

It should be stressed that, in principle, under favorable circumstances, any layered material should be able to form scrolls. Therefore, we should expect new forms of nanoscrolls to be reported in the next years, opening the possibility for exciting novel technological applications. 
\section{Summary}

In summary, nanoscrolls are very unique nanostructures due to their open ended morphology. Such morphology creates the possibility of many different technological applications. However, synthesis difficulties had precluded these structures from being more widely investigated. Recent advances in synthesis techniques had led to a change in this scenario, as interest in nanoscrolls is re-emerging and practical applications are becoming a reality. 

The recent experimental realization of boron nitride nanoscrolls and other scroll materials open new perspectives in the study of these nanostructures. Also, there are many other potential candidates for the formation of novel scrolled structures. In light of this, we can expect in the near future not only significant advances in the production and applications of carbon nanoscrolls, but also the emergence of nanoscrolls from different materials with their own unique properties that can be exploited to be the basis of new applications. We hope the present work can stimulate further studies along these lines.



\section*{Acknowledgement}
D. S. Galvao would like to thank and to acknowledge former group members who worked on the scroll projects, Dr. S. Braga, R. Giro, Profs. V. R. Coluci, S. O. Dantas and S. B. Legoas. The authors would also like to thank Profs. R. H. Baughman, N. M. Pugno, R. Ruralli, and D. Tomanek for many helpful discussions.

\paragraph{Funding:} Work partially funded by Brazilian agencies CNPq, CAPES and FAPESP. The  authors acknowledge  the  Center  for  Computational
Engineering and  Sciences  at  Unicamp  for financial  support  through the FAPESP/CEPID Grant \#2013/08293-7.

\end{document}